# Analyzing phonetic structure of Mandarin using Audacity

**Xu Shizheng**

Computer Science, University of Waterloo

s275xu@uwaterloo.ca

## Abstract

Mandarin Chinese is the official language in China, Taiwan, and Singapore. It is also the main non-official language spoken predominantly at home in Toronto and Vancouver [1]. This article employs the audio software Audacity and leverages theoretical knowledge to conduct a comprehensive analysis of Mandarin Chinese. The study initiates with an overview of the fundamental principles underlying Mandarin pronunciation, aiming to provide insights into its phonetic structure.

## Introduction

The phonological framework of Mandarin Chinese is intricately composed, fundamentally encapsulating three pivotal components that define the structure of its syllables: initials (声母), finals (韵母), and coda (尾音). Figure 1 lists all possible syllable combinations in Mandarin Chinese. Within the realm of Mandarin Chinese phonology, 'initials' are identified as the consonantal sounds that inaugurate a syllable. Progressing through the syllable, 'finals' are characterized by the main vowel or sequence of vowels, which articulate the core auditory essence of the syllable. Following the 'finals,' a 'coda' may emerge, delineating the concluding sound of a syllable that succeeds the primary vowel or vowels. It is imperative to acknowledge that the 'coda' component, while crucial in linguistic analytical discourse, may not always be overtly accentuated in conventional discussions pertaining to Mandarin phonetics. This study aims to dissect these components with precision, shedding light on their roles and interactions within the phonological landscape of Mandarin Chinese.

| Pinyin table | | Initials | | | | | | | | | | | | | | | | | | | | | Pinyin table | |
|---|---|---|---|---|---|---|---|---|---|---|---|---|---|---|---|---|---|---|---|---|---|---|---|---|
| | | (no initial) | b | p | m | f | d | t | n | l | g | k | h | j | q | x | zh | ch | sh | r | z | c | s | | |
| | (no final) | | | | | | | | | | | | | | | | zhi | chi | shi | ri | zi | ci | si | (no final) | |
| | a | a | ba | pa | ma | fa | da | ta | na | la | ga | ka | ha | | | | zha | cha | sha | | za | ca | sa | a | |
| | o | o | bo | po | mo | fo | | | | | | | | | | | | | | | | | | o | |
| | e | e | | | me | | de | te | ne | le | ge | ke | he | | | | zhe | che | she | re | ze | ce | se | e | |
| | ê | | | | | | | | | | | | | | | | | | | | | | | ê | |
| | ai | ai | bai | pai | mai | | dai | tai | nai | lai | gai | kai | hai | | | | zhai | chai | shai | | zai | cai | sai | ai | |
| Group a Finals | ei | ei | bei | pei | mei | fei | dei | | nei | lei | gei | kei | hei | | | | zhei | | shei | | zei | | | ei | Group a Finals |
| | ao | ao | bao | pao | mao | | dao | tao | nao | lao | gao | kao | hao | | | | zhao | chao | shao | rao | zao | cao | sao | ao | |
| | ou | ou | | pou | mou | fou | dou | tou | nou | lou | gou | kou | hou | | | | zhou | chou | shou | rou | zou | cou | sou | ou | |
| | an | an | ban | pan | man | fan | dan | tan | nan | lan | gan | kan | han | | | | zhan | chan | shan | ran | zan | can | san | an | |
| | en | en | ben | pen | men | fen | den | | nen | | gen | ken | hen | | | | zhen | chen | shen | ren | zen | cen | sen | en | |
| | ang | ang | bang | pang | mang | fang | dang | tang | nang | lang | gang | kang | hang | | | | zhang | chang | shang | rang | zang | cang | sang | ang | |
| | eng | eng | beng | peng | meng | feng | deng | teng | neng | leng | geng | keng | heng | | | | zheng | cheng | sheng | reng | zeng | ceng | seng | eng | |
| | er | er | | | | | | | | | | | | | | | | | | | | | | er | |
| | i | yi | bi | pi | mi | | di | ti | ni | li | | | | ji | qi | xi | | | | | | | | i | |
| | ia | ya | | | | | | | | lia | | | | jia | qia | xia | | | | | | | | ia | |
| | io | yo | | | | | | | | | | | | | | | | | | | | | | io | |
| | ie | ye | bie | pie | mie | | die | tie | nie | lie | | | | jie | qie | xie | | | | | | | | ie | |
| | iai | yai | | | | | | | | | | | | | | | | | | | | | | iai | |
| Group i Finals | iao | yao | biao | piao | miao | | diao | tiao | niao | liao | | | | jiao | qiao | xiao | | | | | | | | iao | Group i Finals |
| | iu | you | | | miu | | diu | | niu | liu | | | | jiu | qiu | xiu | | | | | | | | iu | |
| | ian | yan | bian | pian | mian | | dian | tian | nian | lian | | | | jian | qian | xian | | | | | | | | ian | |
| | in | yin | bin | pin | min | | | | nin | lin | | | | jin | qin | xin | | | | | | | | in | |
| | iang | yang | | | | | | | niang | liang | | | | jiang | qiang | xiang | | | | | | | | iang | |
| | ing | ying | bing | ping | ming | | ding | ting | ning | ling | | | | jing | qing | xing | | | | | | | | ing | |
| | u | wu | bu | pu | mu | fu | du | tu | nu | lu | gu | ku | hu | | | | zhu | chu | shu | ru | zu | cu | su | u | |
| | ua | wa | | | | | | | | | gua | kua | hua | | | | zhua | chua | shua | | | | | ua | |
| | uo | wo | | | | | duo | tuo | nuo | luo | guo | kuo | huo | | | | zhuo | chuo | shuo | ruo | zuo | cuo | suo | uo | |
| | uai | wai | | | | | | | | | guai | kuai | huai | | | | zhuai | chuai | shuai | | | | | uai | |
| Group u Finals | ui | wei | | | | | dui | tui | | | gui | kui | hui | | | | zhui | chui | shui | rui | zui | cui | sui | ui | Group u Finals |
| | uan | wan | | | | | duan | tuan | nuan | luan | guan | kuan | huan | | | | zhuan | chuan | shuan | ruan | zuan | cuan | suan | uan | |
| | un | wen | | | | | dun | tun | | lun | gun | kun | hun | | | | zhun | chun | shun | run | zun | cun | sun | un | |
| | uang | wang | | | | | | | | | guang | kuang | huang | | | | zhuang | chuang | shuang | | | | | uang | |
| | ong | weng | | | | | dong | tong | nong | long | gong | kong | hong | | | | zhong | chong | | rong | zong | cong | song | ong | |
| | ü | yu | | | | | | | nü | lü | | | | ju | qu | xu | | | | | | | | ü | |
| | üe | yue | | | | | | | nüe | lüe | | | | jue | que | xue | | | | | | | | üe | |
| Group ü Finals | üan | yuan | | | | | | | | lüan | | | | juan | quan | xuan | | | | | | | | üan | Group ü Finals |
| | ün | yun | | | | | | | | lün | | | | jun | qun | xun | | | | | | | | ün | |
| | iong | yong | | | | | | | | | | | | jiong | qiong | xiong | | | | | | | | iong | |
| Pinyin table | | (no initial) | b | p | m | f | d | t | n | l | g | k | h | j | q | x | zh | ch | sh | r | z | c | s | Pinyin table | |
| | | | | | | | | | | Initials | | | | | | | | | | | | | | | |

Fig 1: Table of Possible Combinations of Chinese Initials and Finals. Retrieved from
*https://en.wikibooks.org/wiki/Chinese_(Mandarin)/Table_of_Initial-Final_Combinations*

In this study, Audacity will serve as the primary tool for phonological examination, with a focus on two principal methods of analysis. Primarily, the investigation will delve into the frequency analysis, shown in Figure 2, to discern the acoustic properties of Mandarin syllables. Additionally, as shown in Figure 3, an Enhanced Autocorrelation (EAC) analysis will be employed to further elucidate the temporal aspects of the speech signal. These methodologies are selected for their proven efficacy in revealing the nuanced details of speech patterns within the domain of phonetics.

| 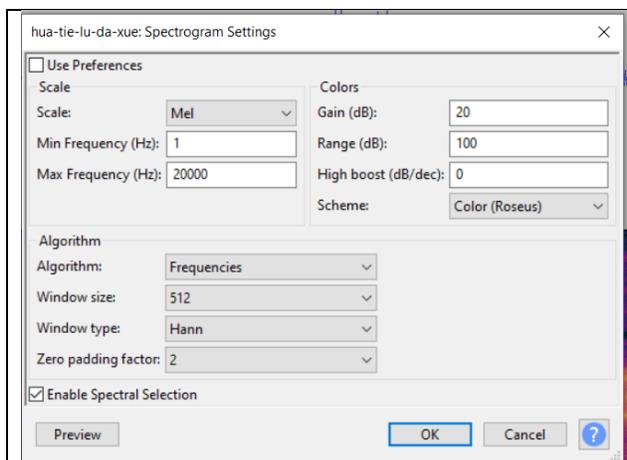 | 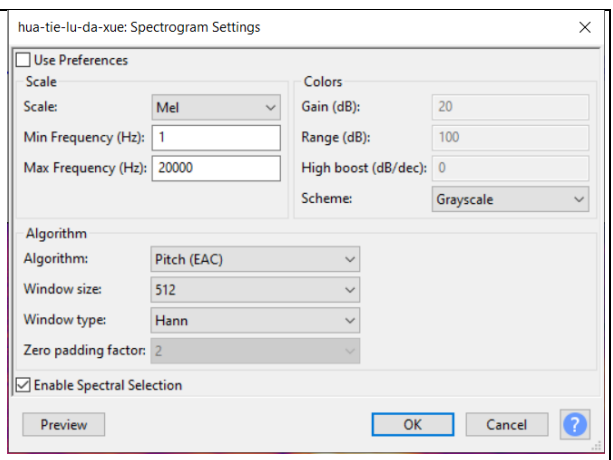 |
|---|---|
| Fig 2: analyze by frequency using Audacity | Fig 3: analyze by EAC using Audacity |

## Overview

In the presented analysis, auditory disparities between the English and Chinese pronunciation spectra of the phrase "University of Waterloo" are elucidated through visual representation. Utilizing state-of-the-art speech synthesis models, namely ElevenLabs [2] for English and Narakeet [3] for Chinese, consistency is maintained across the phonetic investigation. The English denomination, "University of Waterloo," manifests as an unbroken sequence, lacking explicit demarcation between individual words. This continuity is emblematic of English speech patterns. In juxtaposition, the Pinyin rendition of the university's name, "huá tiě lú dà xué"（滑铁卢大学)", exhibits distinct spacing between syllables, mirroring the discrete segmentation intrinsic to Chinese orthography. As shown in Figure 4, an obvious distinction observed is the conspicuous intersyllabic pause in the Chinese enunciation, delineated by a green line within the spectral imagery, a feature absent in the English pronunciation.

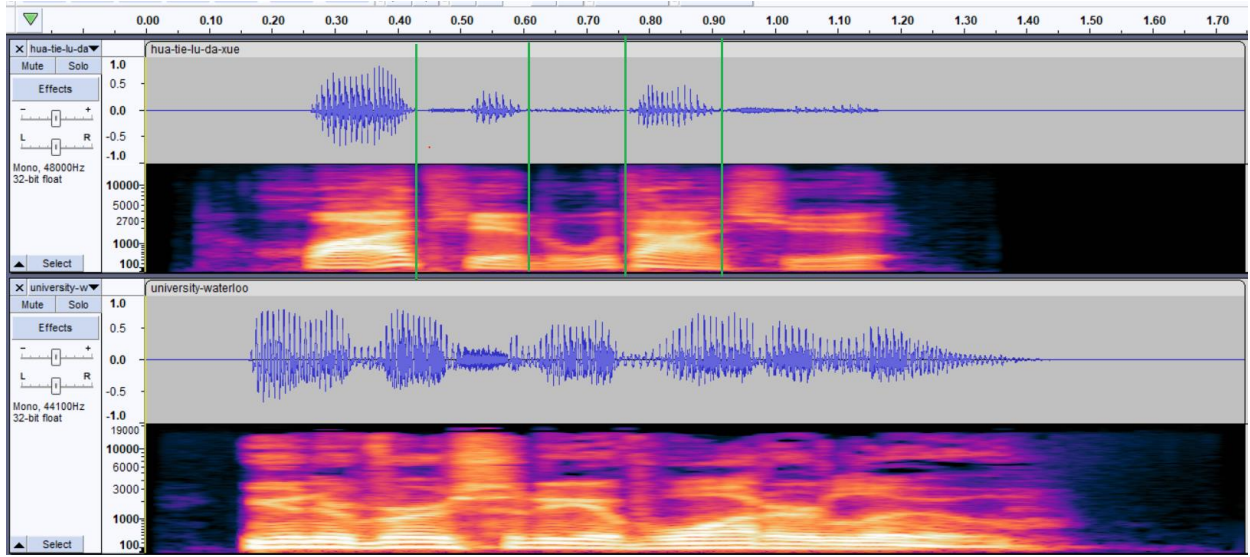

Fig 4: a clear dividing line is seen between each pronunciation in Chinese.

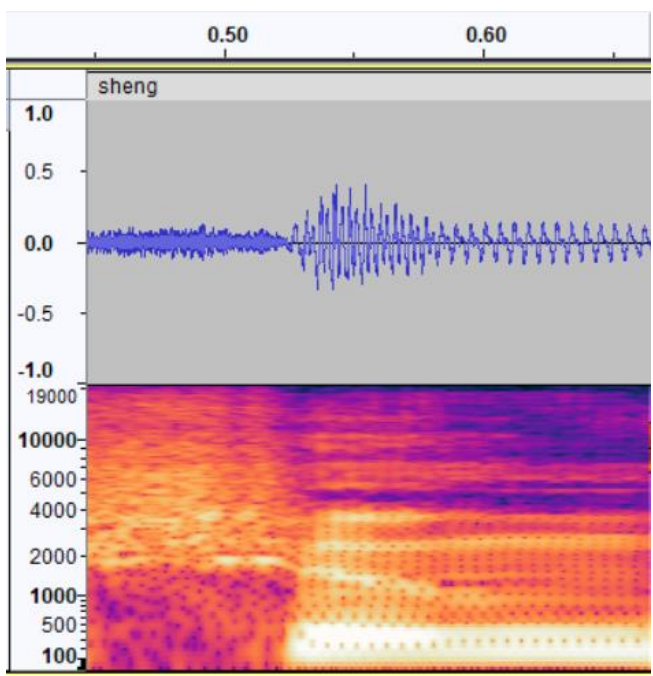

Furthermore, in the upcoming example of "shēng" that we will employ, it is clearly observable that the syllabic structure of Chinese can be distinctly segmented into three main components: the initial consonant (sh), the medial vowel (e), and the final nasal sound (ng). This delineation is markedly evident in a spectrogram.

Fig 5: spectrogram of shēng

## Initials

Within the phonological structure of Mandarin Chinese, the consonant that commences a syllable is designated as the "initial" (声母). This analysis focuses on two exemplary cases: the Mandarin pronunciation of "University Waterloo" (huá tiě lú dà xué), displayed in Figure 6, and the sequence of Chinese numerals from one to ten (yī, èr, sān, sì, wǔ, liù, qī, bā, jiǔ, shí), in Figure 7. Additionally, I will utilize the sound "tā", in Figure 8, to analyze the differences between "d" and "t".

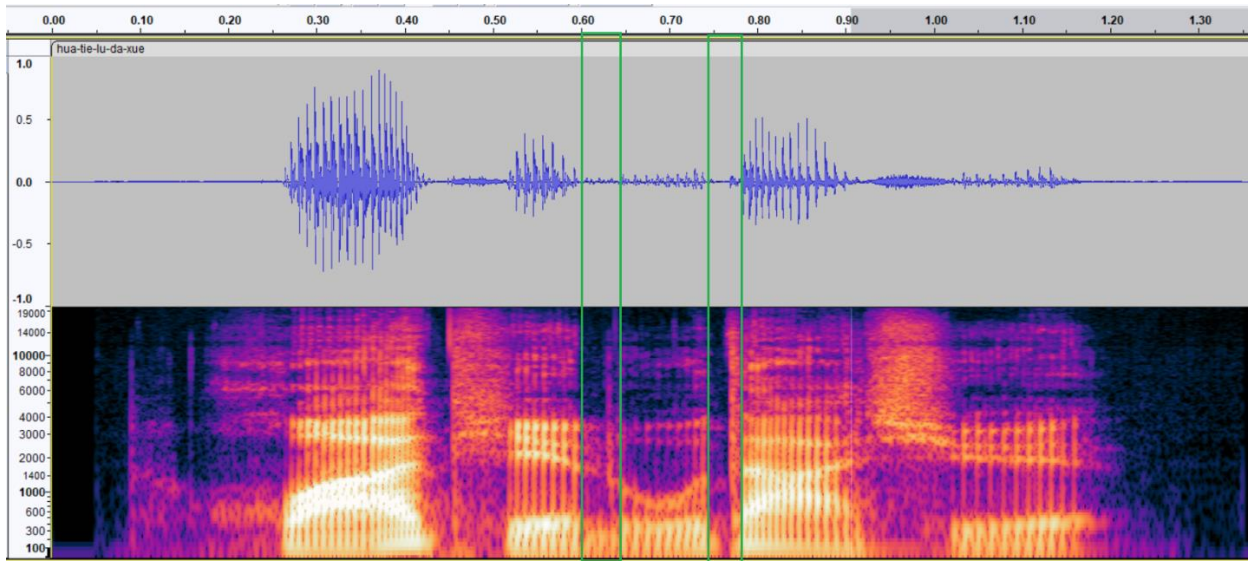

Fig 6: the first green box is 'l', and the second green box is 'd'

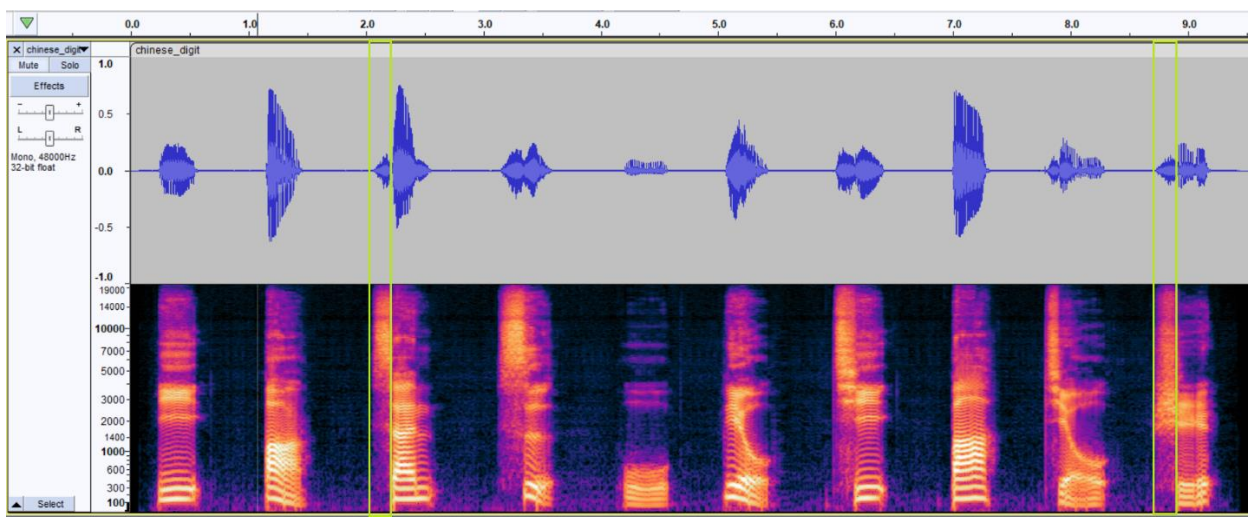

Fig 7: the first yellow box is 's', and the second yellow box is 'sh'

The investigation initiates with scenarios wherein an initial is succeeded by a solitary vowel, exemplified by "lú" and " dà " from the first audio sample, and "sān" and "shí" from the second. These instances provide a foundation for examining the characteristics of short versus long consonants in Mandarin.

Preliminary analysis reveals a dichotomy between short consonants, such as "l" and "d", and their long counterparts, "s" and "sh". The distinction is palpable both in the duration and the spectral composition of these sounds. Notably, the "sh" consonant exhibits a broad spectrum with a pronounced concentration of energy within the higher frequency domain, ranging from 2000 to 8000 Hz. Unlike "s", which manifests a slight preponderance of energy at the uppermost frequencies, "sh" maintains a significant presence in the lower frequency ranges without marked attenuation at higher frequencies.

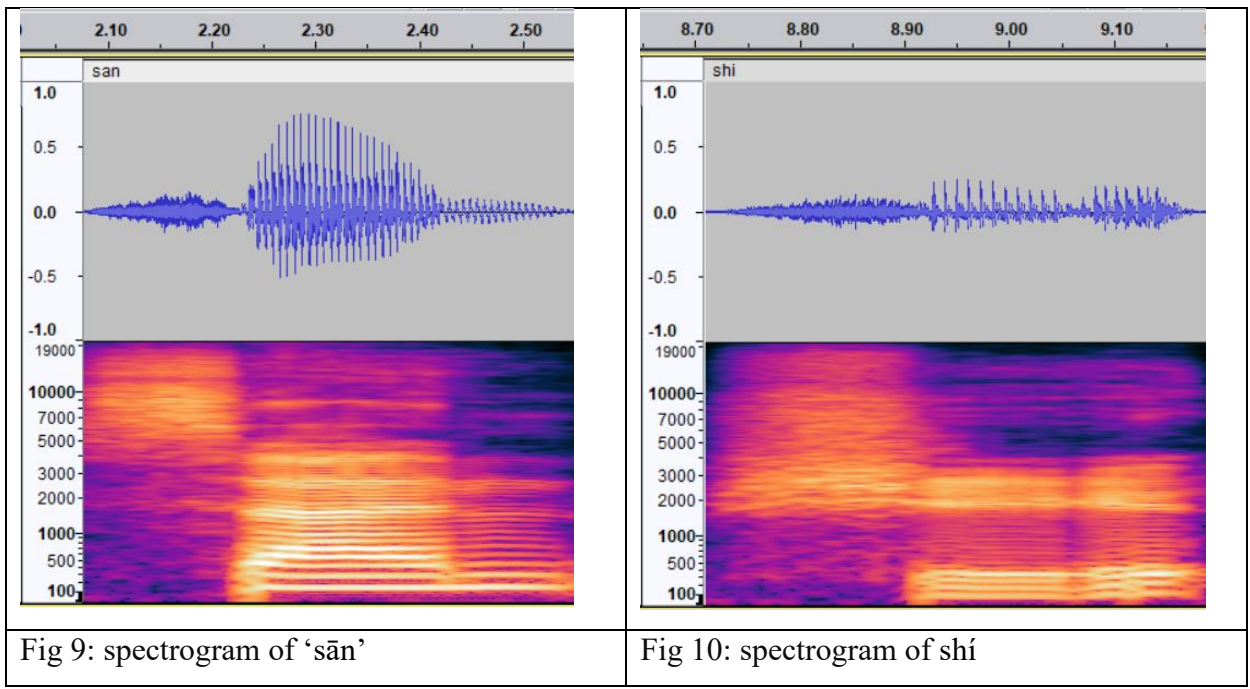

| Fig 9: spectrogram of 'sān' | Fig 10: spectrogram of shí |

Comparatively, the "d" consonant is characterized by a rapid escalation in energy, distinguishing itself from "t" through a pronounced presence in lower frequencies, attributed to the involvement of vocal cord vibrations.

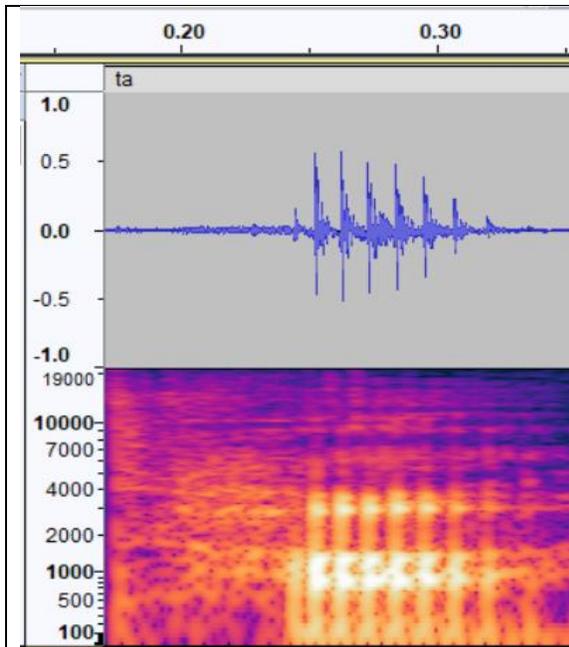

Fig 11: spectrogram of 'tā'

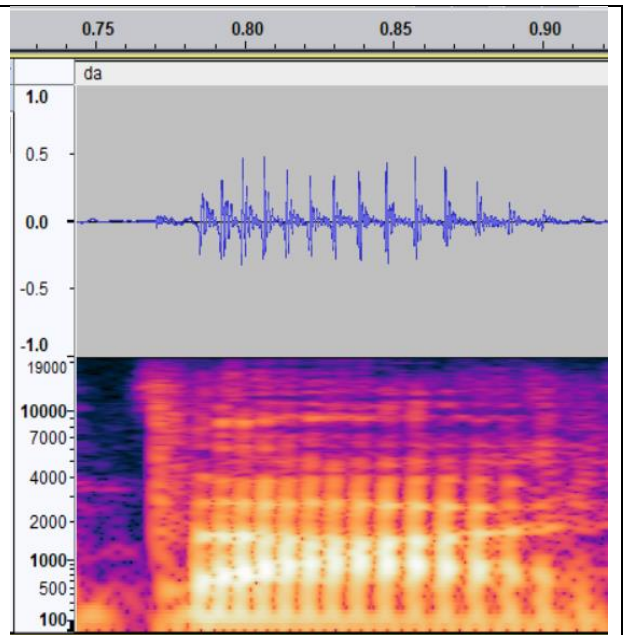

Fig 12: spectrogram of dà

Furthermore, the "l" sound, being voiced, inherently possesses a more substantial low-frequency energy spectrum due to the vibratory actions of the vocal cords. However, the attenuation of energy in the high-frequency spectrum of the "l" sound does not exhibit a pronounced decrease, underscoring the nuanced acoustic profiles of Mandarin initials.

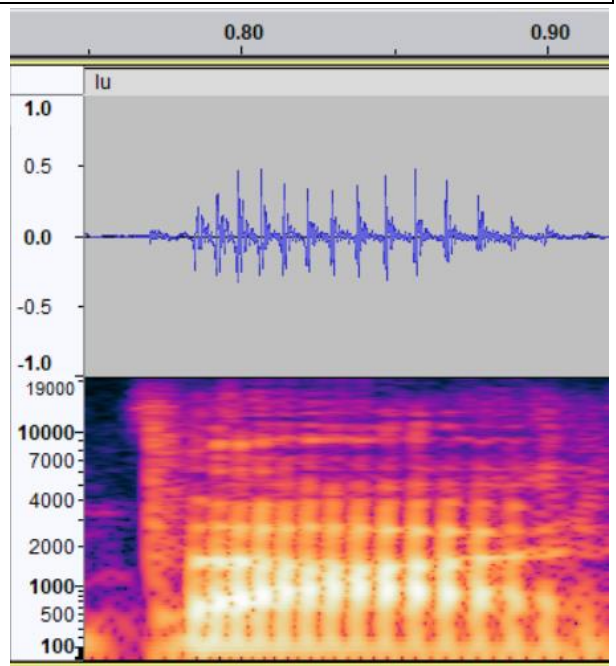

Fig 13: Spectrogram of lú

## Initial – Final Transition

The transition from consonants to vowels warrants particular attention in the study of speech signals, as it exhibits unique spectral characteristics on spectrograms. This section delves into the spectral features of three principal parts of speech signals: consonant, consonant-vowel combination, and pure vowel sections—and analyzes their potential impact on speech processing technologies. See Figure 14 and 15 for how tiě is pronounced, where it demonstrated the transition from t -> ti -> iě.

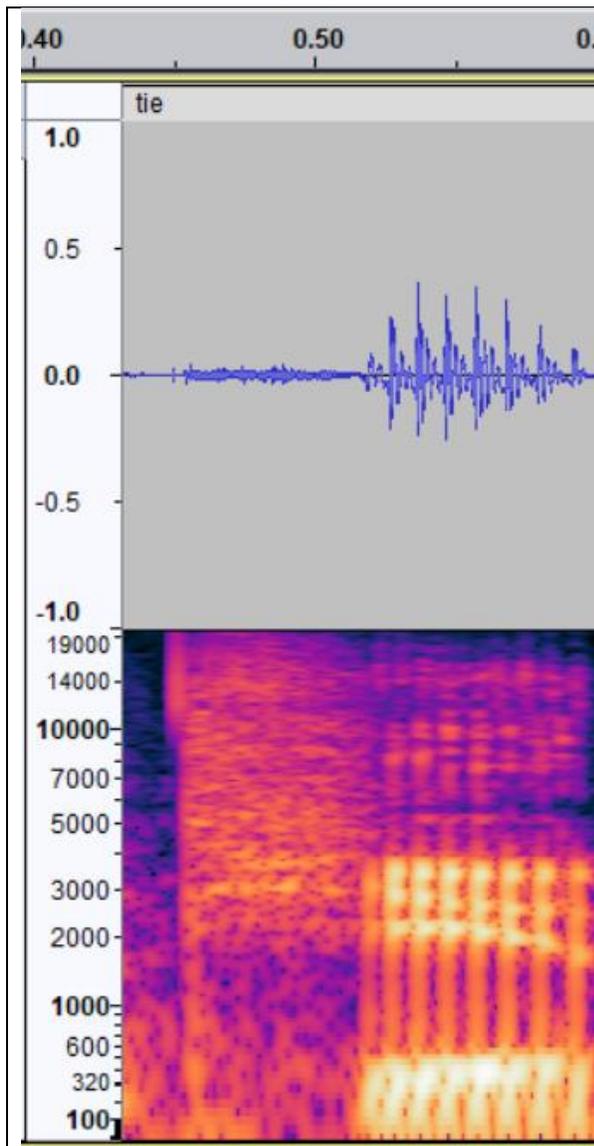 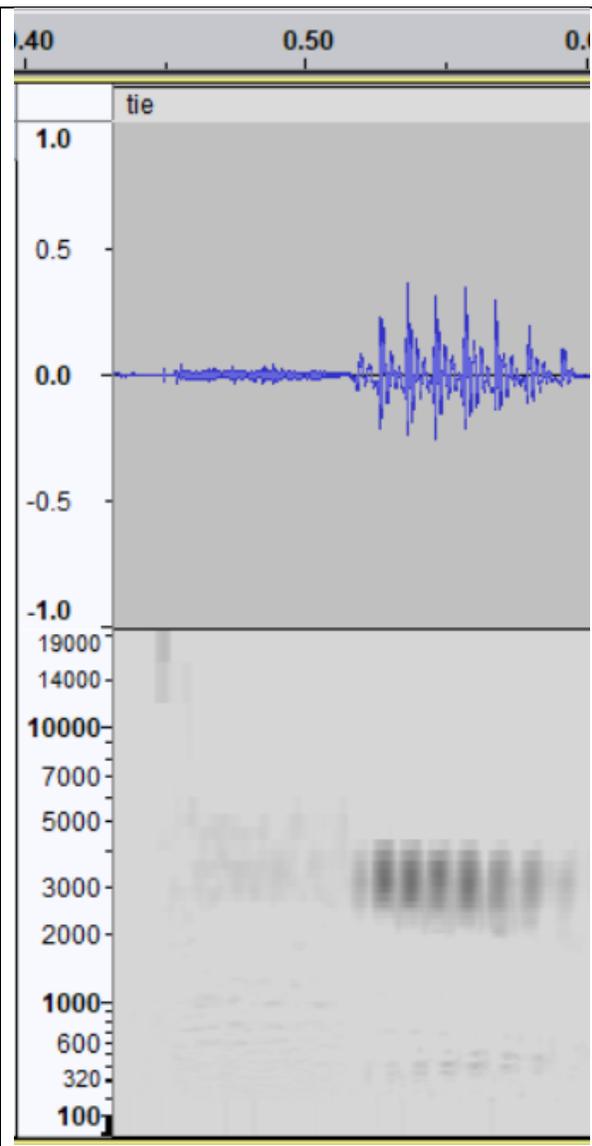

Fig 14: frequency of tiě | Fig 15: EAC of tiě

**Consonants**

The initial segment of the speech signal is characterized by consonant-only sequences. The spectral properties of these sequences are fundamentally influenced by the consonant's manner and place of articulation, as well as the degree of airflow obstruction involved in their production. Consonants are articulated through various restrictions and closures in the vocal tract, resulting in specific spectral patterns. These patterns lack the pronounced formant peaks associated with vowels but are crucial for distinguishing different consonant sounds.

**Vowels**

The final segment predominantly consists of vowels, which may occasionally be followed by codas. This vowel-centric region is marked by distinct formant peaks on the spectrogram, indicative of the vowel sounds' bright frequencies. Vowels are produced with a relatively open vocal tract, allowing for the resonance that generates clear formant structures. These structures are essential for vowel identification and contribute significantly to the intelligibility of speech. The detailed analysis of these formant peaks, which will be elaborated upon in subsequent sections, sheds light on the acoustic nature of vowel sounds.

**Consonant – Vowel Transitions**

The intermediate segment represents a fascinating blend of consonants and vowels, showcasing a distinct spectral profile on the spectrogram. Unlike the consonant-only segment, this part exhibits a higher frequency density, indicative of the presence of vowels. However, it does not display the prominent bright frequencies characteristic of the vowel-dominant segment, highlighting a transitional phase in speech articulation. This segment captures the gradual shift from the articulated consonant sounds to the resonant vowel production, evidencing a smooth phonetic transition that bridges the initial consonant sounds with the subsequent vowel articulations.

## Final Only

Sometimes, there does not exist initials in a word. For example, in the Chinese digits, yī (1), èr (2) and wǔ (5) does not have initials. This can be shown in the spectrogram Figure 16, where there is only one energy distribution.

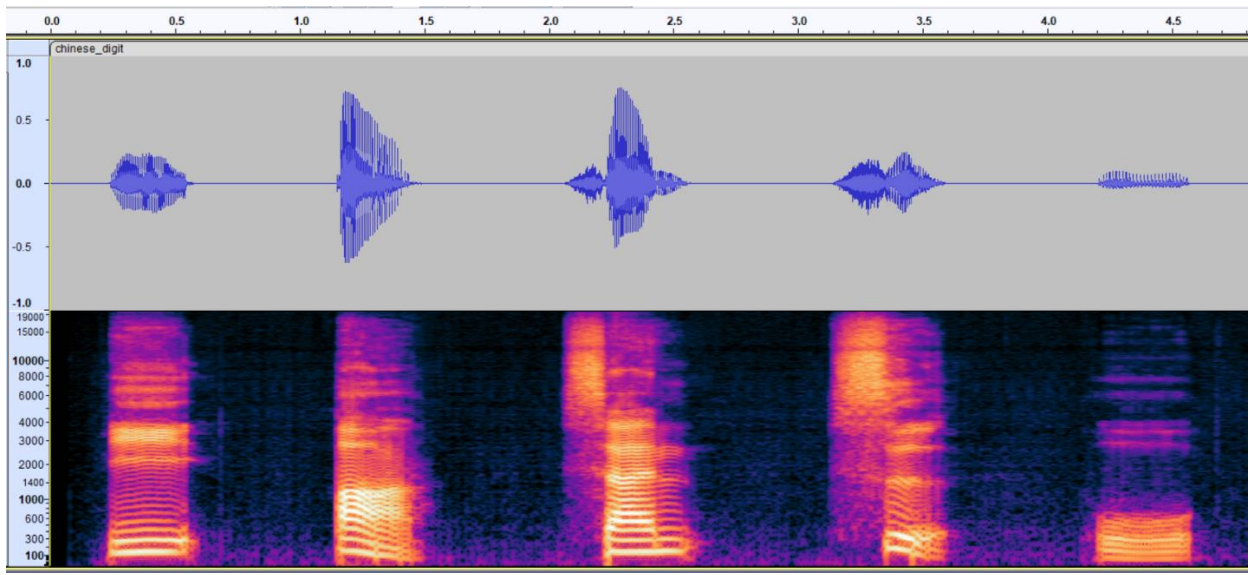

Fig 16: spectrogram of yī – 1, èr – 2 and wǔ – 5

## Finals

In Chinese, "finals" are referred to as "韵母" (yùnmǔ), where "韵" (yùn) signifies the concept of tone. This section will focus on demonstrating how tones vary within the finals. Prior to this, an overview of some basic theories, including the definition of "five-level tone marks," will be provided.

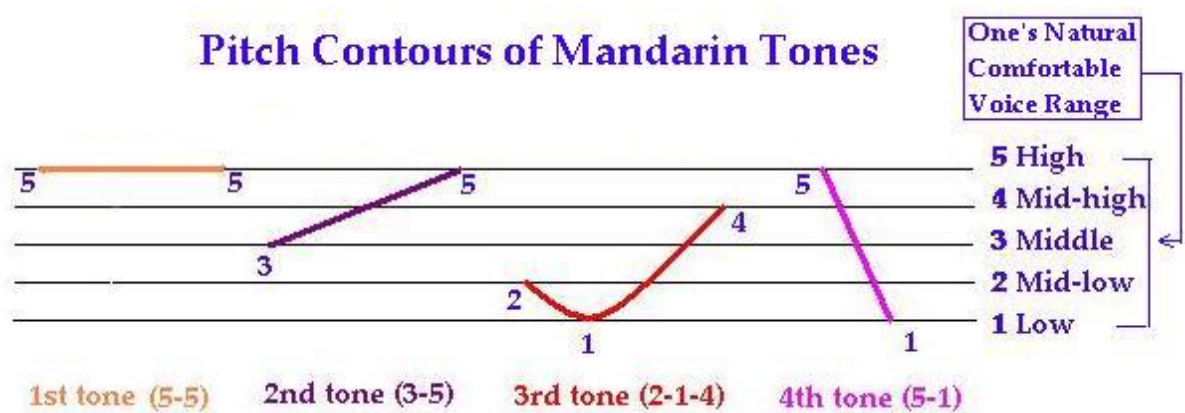

Fig 17: Five Level Tone Marks. Retrieved from
https://web.mit.edu/jinzhang/www/pinyin/tones/index.html

The first Tone (High Level Tone), marked with a macron (ˉ) above the vowel, indicating a high and steady pitch level. The second Tone (Rising Tone), marked with an acute accent (´) above the vowel, indicating the pitch rises from a medium to a high level. The third Tone (Falling-Rising Tone), marked with a caron (ˇ) above the vowel, indicating the pitch first falls then rises. Note that in actual speech, the third tone often manifests as a low pitch, with the rising part frequently omitted. The fourth Tone (Falling Tone), is marked with a grave accent (`) above the vowel, indicating the pitch falls sharply from high to low.

In addition, there exists a neutral Tone (also known as the Fifth Tone or Light Tone), though not marked with any specific tone mark, but its pronunciation is lighter and shorter, with the pitch varying depending on the preceding tone.

In this analysis, "huá tiě lú dà xué" is used again to analyze the tones from the second to the fourth, see Figure 18 and 19.

The first final, "uá," is characterized by an upward trajectory in the spectrogram.

The second final, "iě," demonstrates a slight decrease in the spectrogram.

The third final, "ú," exhibits a rise in tones towards the end, as indicated by the spectrogram.

The fourth final, "à," is marked by a decrease in frequency.

The fifth final, "ué," mirrors the pattern of the first final, showing an increase in the final section of the tone.

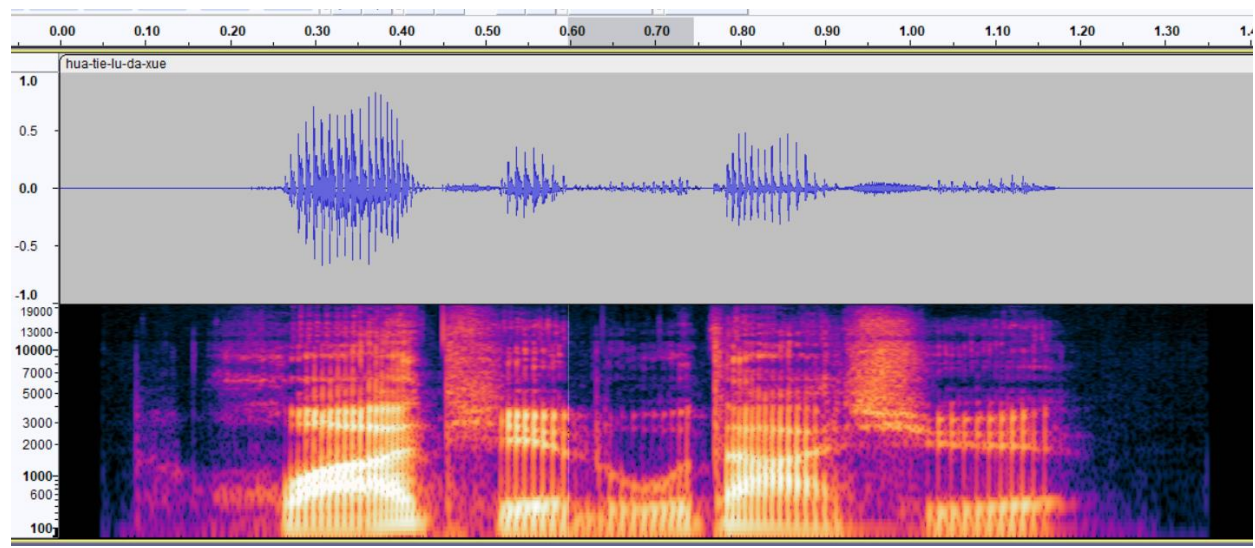

Fig 18: Frequency of huá tiě lú dà xué

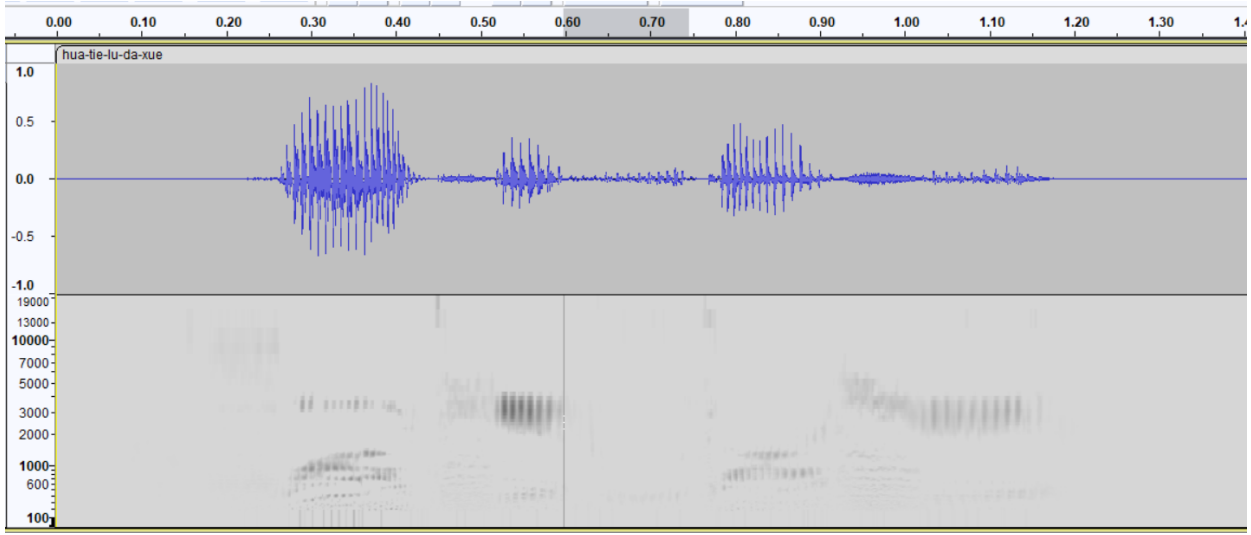

Fig 19: EAC of huá tiě lú dà xué

For the first tone, the previous example sān will be used again. The EAC diagram clearly shows the pitch stays the same (and high) for this sound.

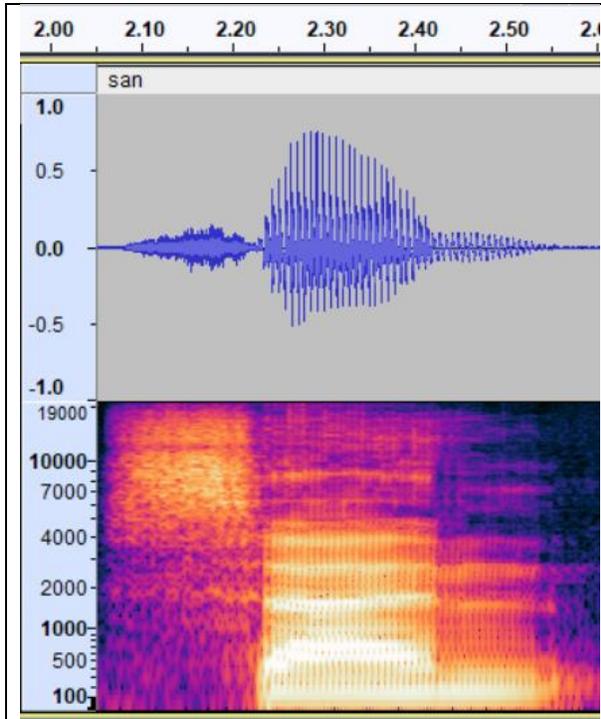

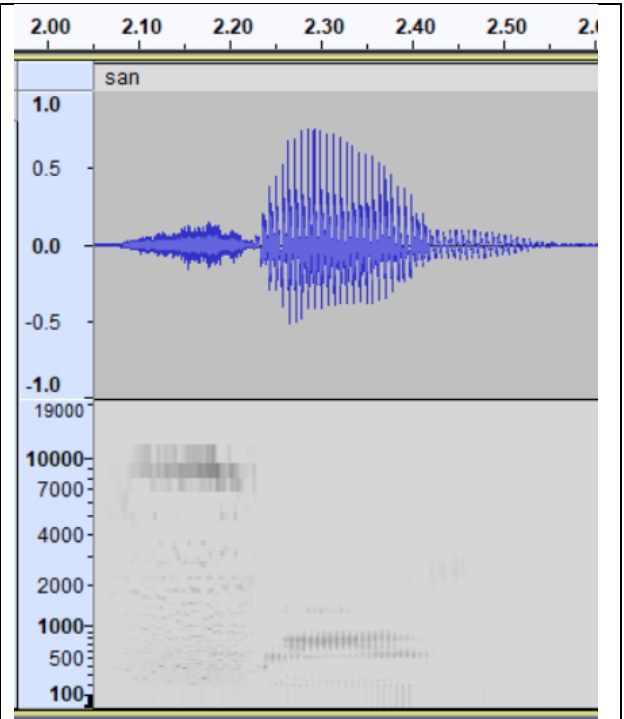

| Fig 20: Frequency of sān | Fig 21: EAC of sān |

## Coda

In Chinese phonetics, there indeed exists the concept of "coda", also known as "final sounds" (尾音), but this term is primarily used in the fields of phonology and linguistics. In the phonetics of the Chinese language, a syllable can generally be divided into the initial (声母), the final (韵母), and the tone, which is inside the final. The final can further be broken down into the initial vowel part and the ending part, where the ending part is referred to as the "final sound", coda. Codas are especially evident in some Chinese dialects that retain features of ancient Chinese pronunciation, such as Cantonese and Min Nan, which may include nasal finals (such as /m/, /n/, /ŋ/) and stop finals (such as /p/, /t/, /k/). Figure 22 shows all possible combinations between vowels and codas.

**Table 5 Cantonese Rhymes**

| V \ final | ∅ | j | ɥ | w | m | n | ŋ | p | t | k |
|---|---|---|---|---|---|---|---|---|---|---|
| iː | iː | | | iːw | iːm | iːn | | iːp | iːt | |
| ɪ | | | | | | | ɪŋ | | | ɪk |
| e | | ej | | | | | | | | |
| ɛː | ɛː | | | ɛw⁵ | | | ɛːŋ | | | ɛːk |
| uː | uː | uːj | | | | uːn | | | uːt | |
| ʊ | | | | | | | ʊŋ | | | ʊk |
| o | | | | ow | | | | | | |
| ɔː | ɔː | ɔːj | | | | ɔːn | ɔːŋ | | ɔːt | ɔːk |
| yː | yː | | | | | yːn | | | yːt | |
| ø | | | øɥ | | | øn | | | øt | |
| œː | œː | | | | | | œːŋ | | | œːk |
| aː | aː | aːj | | aːw | aːm | aːn | aːŋ | aːp | aːt | aːk |
| ʌ | | ʌj | | ʌw | ʌm | ʌn | ʌŋ | ʌp | ʌt | ʌk |

Fig 22: Cantonese Rhymes. (Barrie, 2003)

In Mandarin Chinese, the codas 'n' and 'ng' are retained. The spectrogram illustrated below displays a notable 'tail' following the vowel, with minimal fundamental frequency observed in

EAC. The example of 'sān' was previously discussed, shown in Figure 20 and 21. Additionally, the example of 'shēng' will be presented. The term 'shēng' encompasses multiple meanings in Chinese, one of which is '声', directly relating to the topic of this research, "audio".

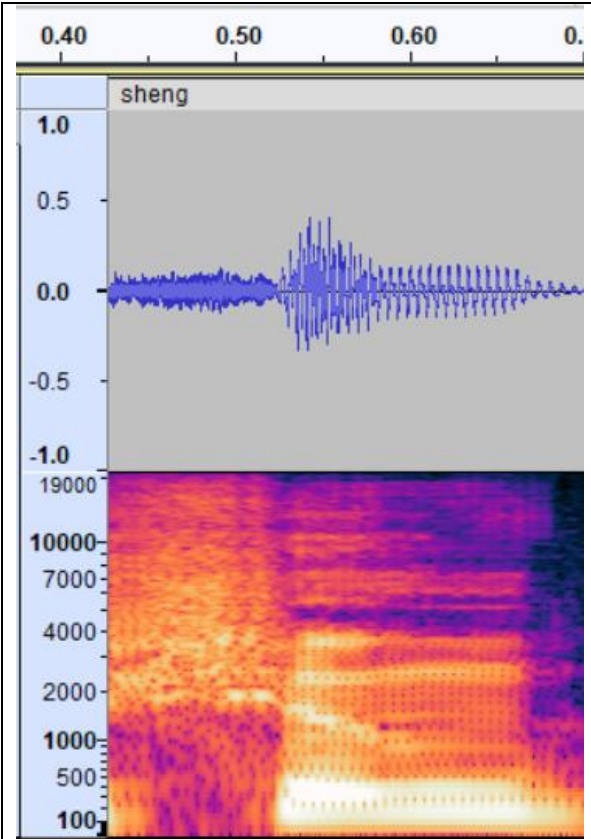 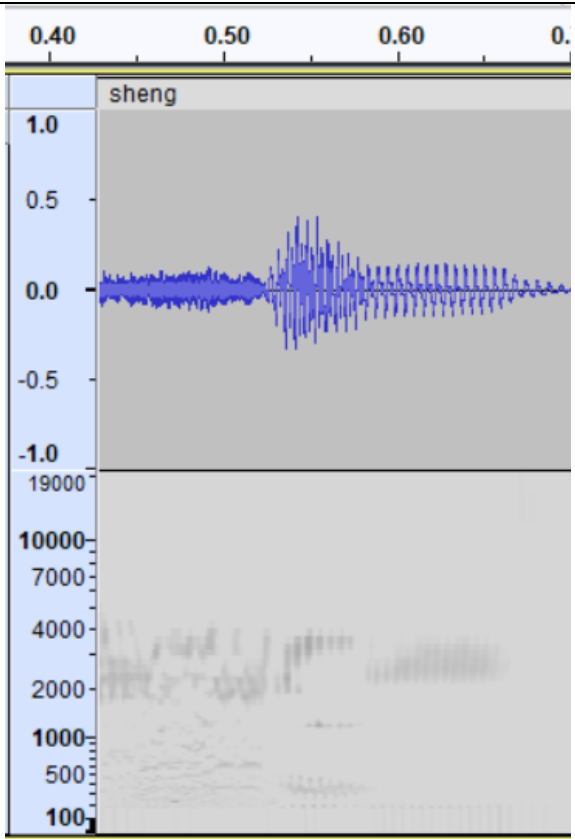

| Fig 22: Frequency of shēng | Fig 23: EAC of shēng |

## Conclusion

In summary, this study has underscored the pronounced segmentation within and between Chinese characters, illustrating the clear delineation of initials, finals, and codas, the distinctive features among different initials, the frequency variations of finals, and the representation of codas in the spectrogram. Particularly, the observation that finals containing multiple vowels result in a noticeable consonant-to-vowel transition in the spectrogram offers valuable insights into the phonetic organization of Chinese. These findings not only enrich our understanding of the Chinese phonetic system but also have potential implications for the fields of linguistic research and speech recognition technology development. Certainly, as the only pictographic writing system in use in the world today, the complexity of Chinese far exceeds the scope of this study.